# Detection of fixed points in spatiotemporal signals by a clustering method

A. Hutt,[1,*] M. Svensén,[1] F. Kruggel,[1] and R. Friedrich[2]
[1]*Max Planck Institute of Cognitive Neuroscience, Stephanstrasse 1a, 04103 Leipzig, Germany*
[2]*Institute for Theoretical Physics, University of Stuttgart, Pfaffenwaldring 57, 70550 Stuttgart, Germany*

We present a method to determine fixed points in spatiotemporal signals. The method combines a clustering algorithm and a nonlinear analysis method fitting temporal dynamics. A 144-dimensional simulated signal, similar to a Kueppers-Lortz instability, is analyzed and its fixed points are reconstructed.

## I. INTRODUCTION

Spatiotemporal signals are obtained in various different research fields [1–3]. Several methods fitting dynamical models in spatiotemporal signals [4–8] cover the data in full time range. We introduce a method to extract underlying fixed point dynamics in discrete time windows by a combination of a clustering algorithm and a nonlinear signal analysis method. The clustering approach represents the main content of the present Rapid Communication. The nonlinear analysis method fits a biorthogonal spatial modes and a system of ordinary differential equations. It validates the introduced approach and represents an important check of the method.

In the following sections, the cluster approach is introduced and applied to a simulated dataset. Subsequently, the nonlinear analysis method aims to determine fixed points of the dataset on the basis of the clustering results. A comparison of the reconstructed and simulated dynamical states verifies the method.

## II. METHOD

### A. The basic idea

We assume that the dynamical system governing the temporal evolution of the system under consideration involves several saddle points (Fig. 1). In order to detect these fixed points, the temporal behavior of the phase trajectories has to be investigated. They approach saddle points along their stable manifolds whereas they leave the vicinity of the fixed points along the unstable manifolds. The phase points accumulate close to the fixed points if the signal is sampled at a constant rate. Thus, the detection of stable manifolds in multidimensional signals can be treated as a recognition problem of point clusters in data space.

In the present paper we use the fuzzy c-means algorithm (FCMA) [9] to detect regions in data space with high density of data points, which are described by its centers and weights for each data point.

### B. The cluster algorithm

An $N$-dimensional spatiotemporal signal can be described by a data vector $\mathbf{q}(t)$, where the component $q_j(t_i)$ represents a data point at time $i$ and detection channel $j$. The clustering algorithm aims at cluster centers $\{\mathbf{k}_k\}$ whose Euclidean distance $d[\mathbf{q}(t_i),\mathbf{k}_k]$ to the datapoints $\mathbf{q}(t_i)$ of the cluster is minimal. These distances are weighted by $u_{ki}^m$ with $0 \leq u_{ki} \leq 1$ which indicate a degree of membership of datapoint $i$ to the cluster $k$. The exponent $m$ is called a fuzzy factor and represents the degree of noise the data is supposed to contain. For $m \to 1$ the data is supposed to contain low noise and one obtains $u_{ki} \to (0,1)$, which is called hard c-means or k-means [10].

Therefore, a cost function

$$J_m = \sum_{k=2}^{c} \sum_{i=1}^{T} (u_{ki})^m d^2(\mathbf{q}(t_i),\mathbf{k}_k) + \sum_{i=1}^{T} \lambda_i \left( \sum_{k=2}^{c} u_{ki} - 1 \right), \quad (2.1)$$

$$d^2(\mathbf{q},\mathbf{k}) = ||(\mathbf{q}-\mathbf{k})||^2,$$

is optimized, where $T$ denotes the number of data points and $c \geq 2$ is the number of clusters. The Lagrange multipliers $\lambda_i$ are introduced to constrain the sum of weights to 1 for each data point. A variation of Eq. (2.1) in respect to $\{u_{ki}\}$ and $\{\mathbf{k}_k\}$ leads to

$$u_{ki} = \left[ \sum_{j=2}^{c} \left( \frac{d^2(\mathbf{q}(t_i),\mathbf{k}_k)}{d^2(\mathbf{q}(t_i),\mathbf{k}_j)} \right)^{1/(m-1)} \right]^{-1}$$

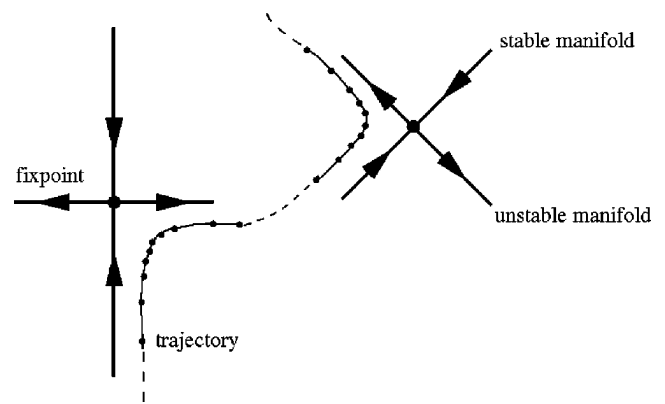

FIG. 1. Trajectory passing saddle points. In this sketch, the dots represent datapoints and the dashed parts of the trajectory represent undetermined transition parts between the fixpoints.

*Electronic address: hutt@cns.mpg.de

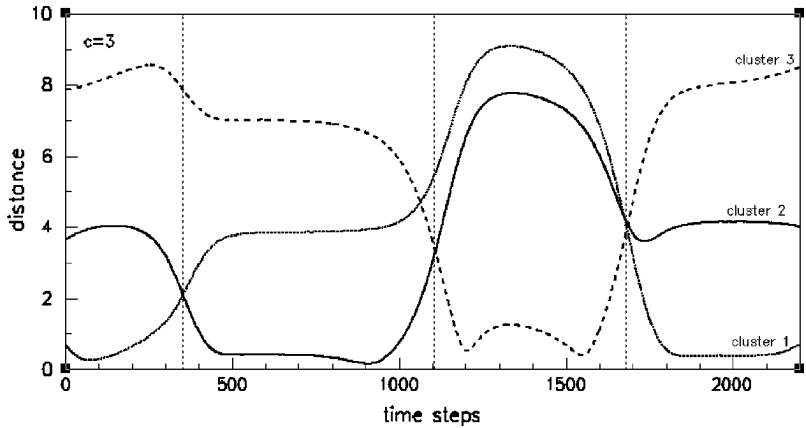

FIG. 2. Distances from each time point to $c=3$ clusters. The time intervals of the states are separated by vertical lines. One recognizes the change between three states and the return of cluster 1.

$$\mathbf{k}_k = \frac{\sum_{i=1}^{T} u_{ki}^m \mathbf{q}(t_i)}{\sum_{i=1}^{T} u_{ki}^m}.$$

The algorithm proceeds by alternatingly reestimating the weights and the cluster means. Typically, it converges within a few tens of iterations.

### III. APPLICATION ON SIMULATED DATA

We applied the cluster algorithm on a 144-dimensional simulated data set. It is generated by a superposition of three spatial modes

$$\mathbf{q}(t) = \sum_{i=1}^{3} A_i(t) \mathbf{v}_i, \tag{3.1}$$

where $A_i(t)$ determines the temporal behavior of spatial modes $\mathbf{v}_i$. We choose three two-dimensional spatial patterns consisting of $12 \times 12$ elements (Fig. 2). The temporal dynamics is determined by the dynamical system

$$\dot{A}_1 = \epsilon A_1 - A_1[A_1^2 + (2+b)A_2^2 + (2-b)A_3^2] + \Gamma(t),$$

$$\dot{A}_2 = \epsilon A_2 - A_2[A_2^2 + (2+b)A_3^2 + (2-b)A_1^1] + \Gamma(t),$$

$$\dot{A}_3 = \epsilon A_3 - A_3[A_3^2 + (2+b)A_1^2 + (2-b)A_2^2] + \Gamma(t),$$

where $\epsilon=1$, $b=2$. $\Gamma(t)$ represents additive noise and follows a uniform deviate with $\Gamma(t) \in [-0.05, \ldots, 0.05]$. The signal is calculated by 2200 integration steps with the initial condition $\mathbf{A}(t=0) = (0.03, 0.2, 0.8)$. Its trajectory passes the saddle points $\mathbf{A}_3^0 = (0,0,1)$, $\mathbf{A}_1^0 = (1,0,0)$, and $\mathbf{A}_2^0 = (0,-1,0)$ in this sequence, and then returns to $\mathbf{A}_3^0$. Thus, these fixed points correspond to the spatial modes $\mathbf{v}_3$, $\mathbf{v}_1$, and $-\mathbf{v}_2$. The dynamical system for the amplitudes $A_i(t)$ arises in a variety of physical systems. We mention the onset of convection in a Rayleigh-Bénard experiment in the presence of rotation, where the amplitude equations describe the so-called Kueppers-Lortz instability [11].

Now we describe the reconstruction of the saddle points from the signal. The method does not use any prior knowledge about the dynamical properties of the signal, a situation which frequently arises in the investigation of real data sets.

### A. Results of clustering

The cluster algorithm was run for 20 iterations ($\|\mathbf{k}_i - \mathbf{k}_{i+1}\| < 10^{-6}$) with a fuzzy factor $m=1.1$ and the number of clusters $c=3$. In Fig. 2, the Euclidean distance from each data point to the different cluster centers is plotted. We consider a data point to be a member of the cluster whose center is closest to the point. If the closest cluster to the data points changes and the data points become members of another cluster, then the trajectory approaches another fixed point. These borders of clusters are marked by vertical dashed lines in Fig. 2. A change of three states where the first occuring cluster returns at the end of the signal is observed. Further investigations on the question of how the results depend on the number of clusters $c$ show consistency with the clustering results for $c=3$. The cluster centers remain stable by increasing $c$ from 2 to 5.

These results allow us a first conclusion. The temporal behavior of the investigated signal can be described by a sequence of three states providing that it is determined from existing stable and unstable manifolds. Therefore, the determined clusters indicate underlying saddle points.

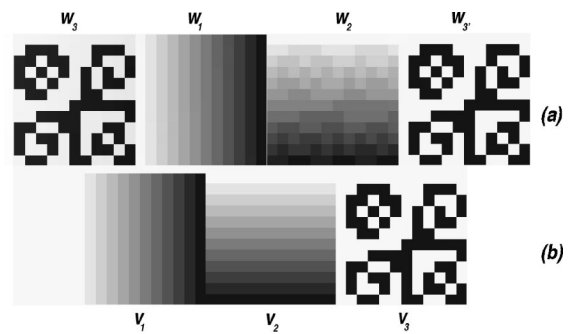

FIG. 3. From left to right: patterns representing (a) reconstructed spatial modes corresponding to the fixed points in time intervals $T_1 = [0 \ldots 350]$, $T_2 = [351 \ldots 1110]$, $T_3 = [1111 \ldots 1680]$, and $T_4 = [1681 \ldots 2200]$ (see Fig. 2), and (b) simulated spatial modes $\mathbf{v}_1, \mathbf{v}_2$, and $\mathbf{v}_3$ corresponding to the fixed points $A_1^0$, $A_2^0$, and $A_3^0$. The necessary accordance of the fixed point in $T_1$ and the returning fixed point in $T_4$ can be recognized in (a).

### B. Application of nonlinear analysis

The clustering method will detect regions of high sample density in data space, but gives on its own no stringent support that these correspond to saddle points. To support this notion, we examined our results by fitting deterministic differential equations in each clustering time window for the case $c=3$.

We used a nonlinear method based on perturbation theory [12]. It fits a polynomial ordinary differential equation system with no constraints for dimension and grade of polynom. Applying it to the signal, one obtains the best fit by a two-dimensional differential equation system with polynoms of second grade for each time window. Now, the stationary solutions of the fitted differential equation systems were used to reconstruct the corresponding spatial modes $\{\mathbf{w}_i\}$ by Eq. (3.1). The reconstructed patterns are shown in Fig. 3(a) in the temporal order of the clustered states. They are in good accordance with the original patterns $\{\mathbf{v}_i\}$, seen in Fig. 3(b).

## IV. CONCLUSION

We introduced a method to detect saddle points in multi-dimensional datasets. The method combines a cluster algorithm presented in this paper and a nonlinear analysis method fitting a model for the dynamics. A high-dimensional simulated dataset was used to illustrate the method. The comparison of reconstructed and original fixed points of the simulated signal showed good accordance. Though we have presented the reconstruction of saddle points, the approach can be generalized on other fixed point dynamics. Signals with underlying unstable limit cycles, unstable tori, or chaotic structures can be investigated since the method just aims at detecting stable and unstable manifolds. This approach may open new ways in the analysis of spatiotemporal signals.

### ACKNOWLEDGMENT

A. Hutt would like to thank A. Hojjatoleslami for fruitful discussions and ideas.